\documentclass[aps,prl,twocolumn,groupedaddress,showpacs]{revtex4}

\usepackage{graphicx}% Include figure files
\usepackage{dcolumn}% Align table columns on decimal point
\usepackage{bm}% bold math
\usepackage{amsmath}

\newcommand{\rpr}{r^{\prime}}

\begin{document}

\preprint{APS/123-QED}

\title{Density fluctuations of a hard-core Bose gas in a one-dimensional lattice}% Force line breaks with \\

\author{C. Ates}
\author{Ch. Moseley}
\author{K. Ziegler}%
\email{Klaus.Ziegler@Physik.Uni-Augsburg.de}
\affiliation{
Institut f\"ur Physik, Universit\"at Ausgburg, D-86135 Augsburg, Germany
}

\date{\today}% It is always \today, today,

\begin{abstract}
For a hard-core Bose gas on a one-dimensional lattice we find characteristic oscillations in the density-density correlation function. 
Their wavelength diverges as the system undergoes a continuous transition from an incommensurate to a Mott insulating phase.
The associated static structure factor vanishes as the Mott insulating phase is approached.
The qualitative picture is unchanged when a weak confinig potential is applied to the system.
\end{abstract}

\maketitle

Recent experiments on one-dimensional (1D) Bose systems 
\cite{moritz03,stoeferle04,paredes04} have opened an exciting new field of
physics. This will provide us with an extended and deeper understanding 
of  the special properties of strongly interacting particles 
at low dimensions. From the theoretical point of view 1D systems are
easier to treat in comparison with two- and three-dimensional
systems but also prevent us from using conventional mean-field
methods.

A well-known fact of one-dimensional many-body physics is that a continuous
hard-core Bose gas is equivalent to a free Fermi gas \cite{girardeau60,lieb63}. 
The role of the hard-core interaction of bosons is played by the statistical
properties of the fermions, related to Pauli's principle. 
This idea can be conveyed to 1D lattice systems. 

In the following we will adopt an approach to the statistics of directed
polymers \cite{forgacs95} in two dimensions. 
The analogy with this classical statistical problem is based on the observation
that the world lines of a grand-canonical ensemble of bosons are equivalent to directed
polymers, random walks or fluctuating flux lines \cite{nelson88,ziegler89}.
This can be formally expressed by the fact that the partition function $Z$ of the
grand-canonical ensemble of these systems is identical. 

For directed polymers in two dimensions it was shown that $Z$ can be expressed as a determinant \cite{forgacs95}.
Thus the partition function of hard-core bosons in $d=1$ reads
\begin{equation}
Z=\det {\bf R} \quad ,
\end{equation}
where ${\bf R}$ is diagonal with respect to the Matsubara frequency
$\omega$
\begin{eqnarray}
{\bf R}(\omega) &=&
(e^{i\omega}-\zeta^{-1})\sigma_0 \nonumber \\
& &{}+\frac{J}{2}(1 + e^{i\omega}+\hat{T}^{-1}+e^{i\omega}\hat{T})\sigma_1
\nonumber \\
& &{}-i\frac{J}{2}(1 - e^{i\omega}+\hat{T}^{-1}-e^{i\omega}\hat{T})\sigma_2
.
\end{eqnarray}
$\hat{T}$ is the shift operator along the 1D lattice 
($\hat{T}f(r)=f(r+1)$), and the $\sigma_j$ are the Pauli matrices
\[
\sigma_0=\left(\begin{array}{cc}1&0\\0&1\end{array}\right) \; ,\;
\sigma_1=\left(\begin{array}{cc}0&1\\1&0\end{array}\right) \; ,\;
\sigma_2=\left(\begin{array}{cc}0&-i\\i&0\end{array}\right) \, .
\]
This model describes the tunneling of bosons with rate $J\ge0$ between 
nearest neighbors, expressed by the shift operator $\hat{T}$ and its
inverse $\hat{T}^{-1}$.
$J$ is measured in units of the lattice potential.
The $2\times 2$ structure arises from the fact that particle exchange between neighboring
sites through simultaneous tunneling is prohibited.
This reduces the translational symmetry to sublattices with every second site,
where $r$ is the position on a sublattice.
$\zeta>0$ is the fugacity that controls the density of bosons in the system.
The fugacity is not directly accessible in the experiment but only indirectly
through
the density $n(\zeta,J)$. Therefore, physical quantities should be measured
as functions of $n$ and $J$.
An additional potential, superimposed on the optical lattice, is described by a
space-dependent fugacity $\zeta_r$. 

Physical quantities can be derived from the matrix ${\bf R}$. 
For instance, if we are interested in properties at zero temperature we can use
the integral with respect to the Matsubara frequency 
\begin{equation}
G_{r,r'}=\int_0^{2\pi}({\bf R}^{-1})_{r,r'}\frac{d\omega}{2\pi} \label{G}
\end{equation}
to evaluate the local density of bosons as
\begin{equation}
n_r = 1+ \zeta^{-1}_rG_{r,r} \quad .\label{TD}
\end{equation}
It should be noticed that $G_{r,r'}$ is not the Green's function for the propagation
 of an individual boson in the system but a correlation function between two bosons. 
Therefore, it is not possible to evaluate the momentum distribution of bosons
\cite{paredes04,girardeau60,lieb63,olshanii98,gangardt03,papenbrock03,
girardeau04,gangardt04,kollath04} from this expression. On the other hand, the
evaluation of the density and the density-density correlation function becomes a simple task with 
the matrix $G_{r,r'}$ \cite{forgacs95}. 

\begin{figure}
\centering
\includegraphics[width=0.45\textwidth]{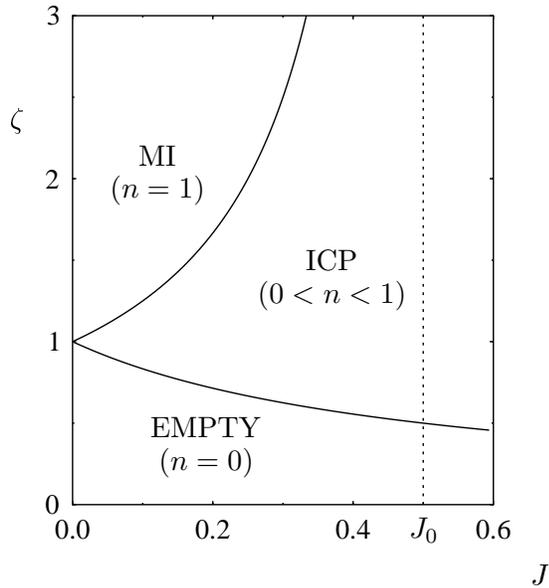}
\caption{The zero-temperature phase diagram of the model shows three phases: a
Mott insulator (MI), an incommensurate (ICP) and an empty phase. For $J>J_0=1/2$
there is no MI.} \label{PhaseDiag}
\end{figure}

\begin{figure*}
\includegraphics[width=0.45\textwidth]{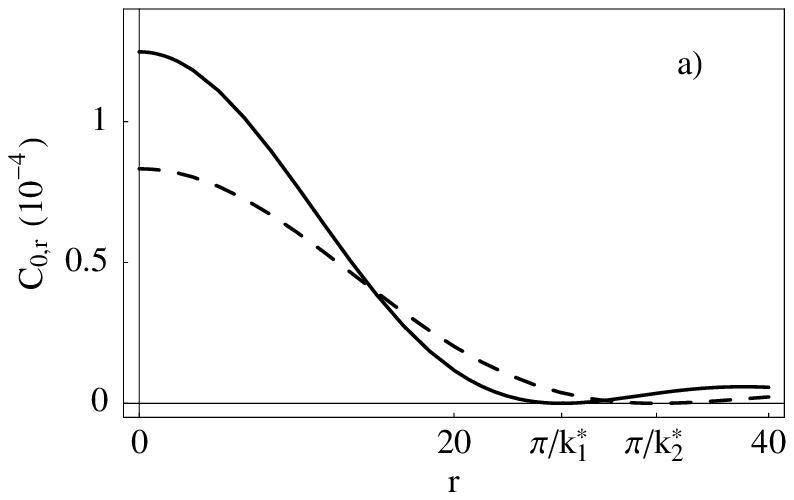}
\includegraphics[width=0.45\textwidth]{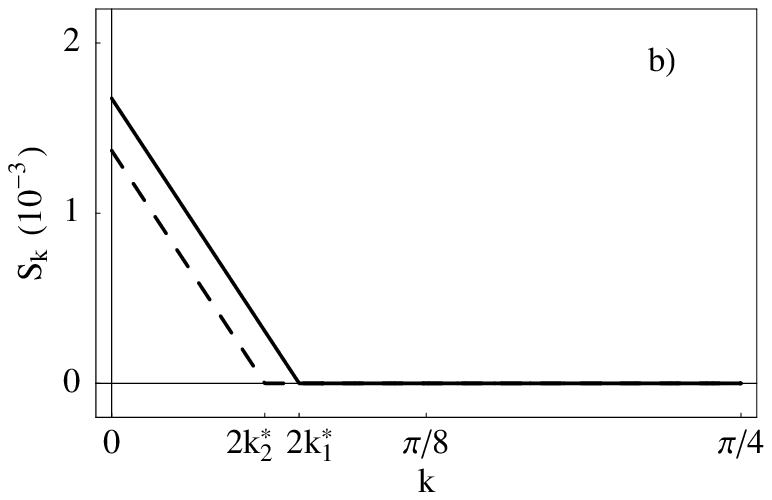}
\includegraphics[width=0.45\textwidth]{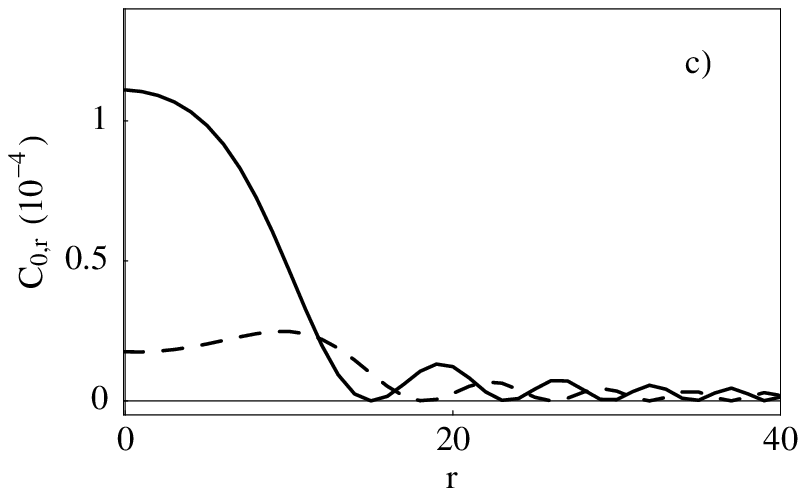}
\includegraphics[width=0.45\textwidth]{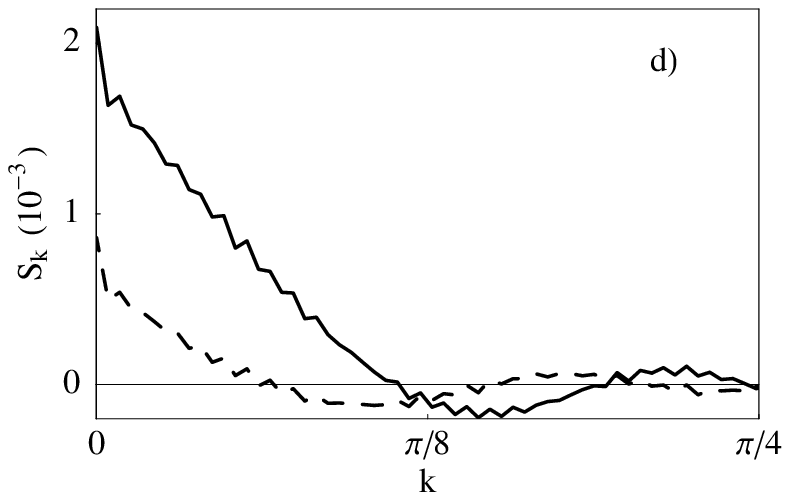}
\caption{Correlation function of density fluctuations $C_{0,r}$ and static structure
factor $S_k$ for different tunneling rates $J$. 
First row: constant background ($\zeta^{-1}=0.3$, $\Omega=0$).
Second row: parabolic background ($\zeta^{-1}=0.3$, $\Omega=3\cdot 10^{-5}$).
Tunneling rates: $J_1=0.3506$ (solid lines) and $J_2=0.3504$ (dashed lines)}\label{CFSFfig}
\end{figure*}

\begin{figure}
\includegraphics[width=0.45\textwidth]{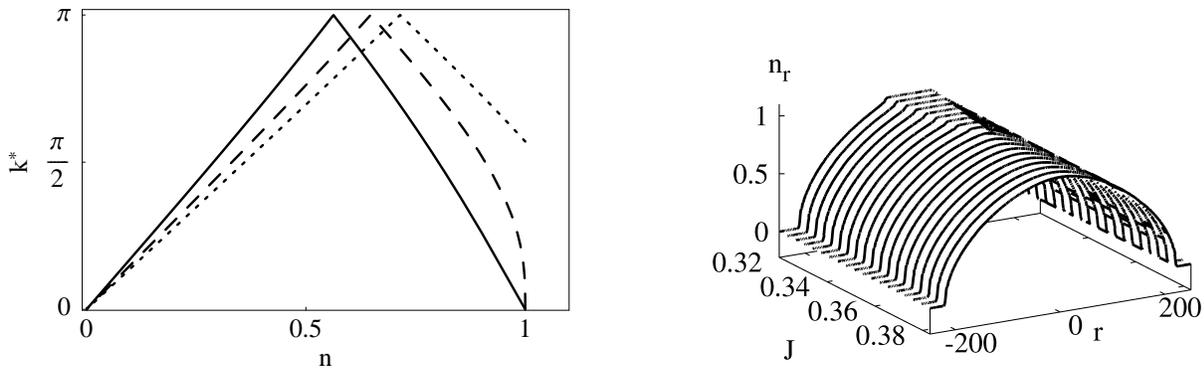}
\caption{Characteristic wave vector $k^{\star}$ as a function of the density in
the translational-invariant case. Curves are plotted for different values of the
tunneling rate: $J=0.2<J_0$ (solid), $J=0.5=J_0$ (dashed), $J=0.8>J_0$ (dotted).} \label{kstarn}
\end{figure}

\begin{figure}
\includegraphics[width=0.45\textwidth]{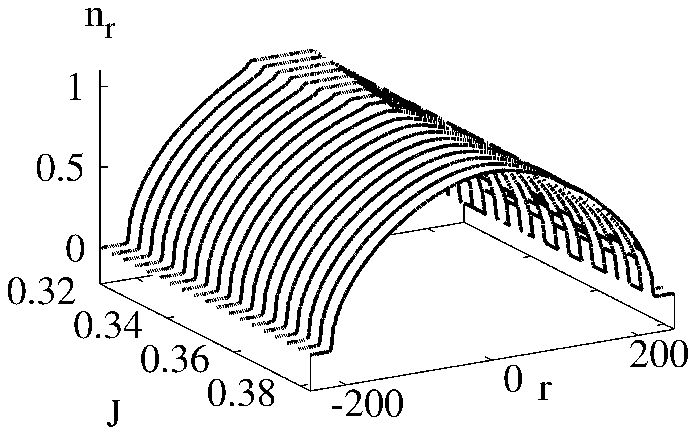}
\caption{Local particle density for parabolic background potential
($\zeta^{-1}=0.3, \Omega=3\cdot 10^{-5} $).
Development of a Mott plateau in the center of the trap ($r=0$) as the
 tunneling rate $J$ is decreased below $J_{\text{P}} \approx 0.35$.} \label{localPD}
\end{figure}

Another interesting quantity is the correlation function of the density fluctuations
\begin{eqnarray}
C_{r,\rpr} &\equiv& -\langle\left(n_r -\langle n_r \rangle  \right) \left(
n_{\rpr} - \langle n_{\rpr} \rangle  \right)\rangle \nonumber\\ 
&=& \left(\zeta_r\zeta_{\rpr}\right)^{-1}
G_{r,\rpr}G_{\rpr,r} \label{KF}
\end{eqnarray}
from which the static structure factor as a function of the momentum $k$
can be obtained by Fourier transformation:
\begin{equation}
S_{k} = \frac{1}{N} \sum_{r} C_{0,r} e^{-ikr} \quad . \label{SF}
\end{equation}
$N$ is the number of sites of the 1D lattice.

The local density $n_r$ and the correlation of the density fluctuations can be
directly measured in an experiment \cite{stamper00}. 
This motivates the following study of these quantities for a translational-invariant
system as well as for a system with a weak parabolic potential. 
We will compare the results in the incommensurate regime near the transition to the Mott insulator and discuss their characteristic properties.   

{\it (i) Translational-invariant case.}
For constant fugacity $\zeta_r\equiv\zeta$ the system has translational symmetry
on the sublattices, and $G_{r,\rpr} $ can be calculated analytically \cite{forgacs95}.

Figure \ref{PhaseDiag} shows the zero-temperature phase diagram of the model.
The particle density is constant in space.
Three phases can be identified: an empty phase with $n=0$ for
$\zeta<1/(1+2J)$, a Mott insulator (MI) with $n=1$ for $\zeta>1/(1-2J)$ and
$J<1/2$, and an incommensurate phase (ICP).
For $J>J_0=1/2$ the system exhibits no MI phase.
The density in the ICP can be calculated from Eq. (\ref{TD}) and gives
\begin{equation}
n = 1 - \frac{1}{2\pi} 
\left[\tilde{k}\mp\left( k^{\star} - \pi\right) \right]\quad ,
\end{equation}
where $\mp$ correspond to the cases $\zeta >1$ and $\zeta<1$, respectively,
 and $\tilde{k}$, $k^{\star}$ are given by
\begin{eqnarray}
\tilde{k} &=& \arccos \left(1 - \frac{(1+\zeta^{-1})^2}{2(\zeta^{-1}+J^2)}
\right) \; ,\\
k^{\star} &=& \arccos\left(\frac{(1-\zeta^{-1})^2}{2J^2}-1\right) \; . 
\end{eqnarray}
The transition from the intermediate to the Mott insulating phase at the
critical fugacity $\zeta_{\text{c}}=1/(1-2J)$ is continuous.

To investigate the behavior of the system near the Mott transition we calculate
the correlations of density fluctuations asymptotically from Eq. (\ref{KF}) as
\begin{equation}
C_{0,r} \sim \left( \frac{\sin\left(k^{\star}r \right)}{\zeta r}\right)^2.
\end{equation}
After a Fourier transformation we obtain the static structure factor as
\begin{equation}
{\renewcommand{\arraystretch}{1.5}
S_{k} \sim \left\{ \begin{array}{c@{\quad :\quad}l}
\frac{\pi}{2\zeta^2} \left(k^{\star} - \frac{1}{2}k \right) & 0 < k <
2k^{\star} \\
\frac{\pi}{2\zeta^2} \left(k^{\star} + \frac{1}{2}k \right) & 0 > -k >
-2k^{\star} \\
0 & |k| > 2k^{\star}
\end{array}
\right.} \, .
\end{equation}
These quantities are shown in Fig. \ref{CFSFfig} a,b for two values of the
tunneling rate $J$ in the ICP.
They vanish as the MI phase is reached due to the fact that the MI
exhibits no density fluctuations.
The correlation function of the density fluctuations shows significant oscillations
in the ICP.
Their wavelength $\lambda=2\pi/k^{\star}$ determines the characteristic length
 scale for density fluctuations and diverges as the Mott transition is approached.
$S_k$ is nonzero in the interval $[-2k^{\star},2k^{\star}]$ and falls off
linearly to both sides from $k=0$. 
The slope of the static structure factor does not depend on the tunneling rate.

The dependence of $k^{\star}$ on the density for varying tunneling rates is depicted in Fig. \ref{kstarn}.
At low densities $k^{\star}$ increases with increasing $n$ until it reaches its
maximal value of $\pi$ at a certain density, where it shows a cusp for $J<\infty$.
For $J\le J_0$ the system can undergo a Mott transition, where $k^{\star}$ vanishes at $n=1$.
Otherwise it stays nonzero.

Note that these features of hard-core bosons in 1D differ fundamentally from
those of an ideal Bose gas in 1D. 
There $C_{0,r}$ decays exponentially and its oscillations show no characteristic
behavior.

{\it (ii) Parabolic background potential.}
A parabolic potential can be expressed as a spatially varying fugacity
$\zeta^{-1}_r=\zeta^{-1}+\Omega r^2$, where $\Omega$ determines the strength
of the potential. 
In this case $G_{r,\rpr}$ cannot be evaluated simply by a Fourier transformation,
since the translational invariance is broken.
We have calculated the local particle density, the correlations of the density
fluctuations and the static structure factor by inverting the matrix ${\bf R}$
 numerically on a lattice with $N=500$ sites.

The local particle density $n_r$ is shown in Fig. \ref{localPD} for different
values of the tunneling rate $J$.
The density is symmetric around the minimum of the parabolic potential at 
$r=0$ with a maximum at the center \cite{dunjko01}. 
It is suppressed as the potential becomes larger with increasing distance
from the center of the trap.
As the tunneling rate is decreased the distribution of the particles along the
lattice becomes narrower and the density is shifted upwards.
When $J$ reaches some value $J_{\text{P}}$, we observe a region with local  particle density $n_r=1$ developing symmetrically around $r=0$.

To investigate the development of this plateau we have evaluated the correlations of the density fluctuations $C_{0,r}$ together with the associated static structure factor.
These quantities are depicted in Fig. \ref{CFSFfig} c,d for two values of $J \gtrsim J_{\text{P}}$.
The correlation function of the density fluctuations exhibits oscillations that
do not have a unique wavelength and $S_k$ does not show a sharp cut-off.
However, the properties are qualitatively the same as in the
translational-invariant case.
Both quantities vanish when $J_{\text{P}}$ is reached, owing to the fact that
there are no density fluctuations within the plateau.
The characteristic length scales become larger as the Mott plateau is approached.
Close to $J_{\text{P}}$ we observe clear indications for the developing Mott
plateau \cite{bergkvist04}.
The correlations of the density fluctuations are suppressed around the center of
the trap leading to a local minimum of $C_{0,r}$ at $r=0$.
This is accompanied by a decrease of the slope of $S_k$.

{\it Conclusion.} For the 1D strongly interacting Bose gas in an
optical lattice we have identified characteristic oscillations of the 
density-density correlation function with length $\lambda$. This can be used
as a measure for the
distance of the system from the MI state: the length $\lambda$ diverges
in units of the lattice spacing 
with the density $n$ as $1/(1-n)$ when we approach the MI. This phenomenon is
related to the behavior of the static structure factor $S_k$ with
characteristic wavevector $k^{\star}=2\pi/\lambda$.  $S_k$ is linear and
vanishes for $|k|>k^{\star}$.
$k^{\star}$ itself vanishes continuously as the MI is approached and $S_k=0$ in the MI phase. This behavior also survives qualitatively if a weak
parabolic potential is applied to the interacting Bose gas. In particular,
the static structure 
factor is strongly suppressed if a large fraction of the Bose gas is in the MI state. 

\begin{acknowledgments}
We are grateful for inspiring discussions with G. Shlyapnikov and M. Girardeau. 
This research was supported in part by the National Science Foundation under Grant No. PHY99-07949.
\end{acknowledgments}


\begin{thebibliography}{99}
\bibitem{moritz03}
H. Moritz et al., cond-mat/0307607
\bibitem{stoeferle04}
T. St\"oferle et al., Phys. Rev. Lett. {\bf 92}, 130403 (2004)
\bibitem{paredes04}
B. Paredes et al., Nature (London) {\bf 429}, 277-281 (2004)
\bibitem{girardeau60}
M. Girardeau, J. Math.. Phys. {\bf 1}, 516 (1960)
\bibitem{lieb63}
E. H. Lieb and W. Liniger, Phys. Rev. {\bf 130}, 1605 (1963)
\bibitem{forgacs95}
G. Forgacs and K. Ziegler, Europhys. Lett. {\bf 29}, 705 (1995)
\bibitem{nelson88}
D. R. Nelson, Phys. Rev. Lett. {\bf 60}, 1973 (1988) 
\bibitem{ziegler89}
K. Ziegler, Europhys. Lett. {\bf 9}, 277 (1989)
\bibitem{olshanii98}
M. Olshanii, Phys. Rev. Lett. {\bf 81}, 938 (1998)
\bibitem{gangardt03}
D. M. Gangardt and G. V. Shlyapnikov, Phys. Rev. Lett. {\bf 90}, 010401 (2003)
\bibitem{papenbrock03}
T. Papenbrock, Phys. Rev. A {\bf 67}, 041601(R), (2003)
\bibitem{girardeau04}
M. Girardeau, H. Nguyen and M. Olshanii, cond-mat/0403721
\bibitem{gangardt04}
D. M. Gangardt, cond-mat/0404104
\bibitem{kollath04}
C. Kollath et al., Phys. Rev. A {\bf 69}, 031601(R) (2004)
\bibitem{stamper00}
D. M. Stamper-Kurn and W. Ketterle, cond-mat/0005001
\bibitem{dunjko01}
V. Dunjko, V. Lorent and M. Olshanii, Phys. Rev. Lett. {\bf 86}, 5413 (2001)
\bibitem{bergkvist04}
S. Bergkvist, P. Henelius and A. Rosengren, cond-mat/0404395
\end{thebibliography}
\end{document}